\documentclass{icrc2009}

\usepackage{graphicx}   
\usepackage[caption=false]{caption}    
\usepackage[font=footnotesize]{subfig} 
\usepackage{fixltx2e}
\usepackage{url}

\newcommand{\shorttitle}[1]%
{\markboth{Proceedings of the 31\MakeLowercase{$^{st}$} ICRC, {\L}\'{o}d\'{z} 2009}{#1} }
\newcommand{\etal}{\MakeLowercase{\textit{et al. }}} 

\begin{document}
\title{Charge Calibration of the ANTARES high energy neutrino telescope.}

\author{\IEEEauthorblockN{Bruny Baret\IEEEauthorrefmark{1} 
			  on behalf of the ANTARES Collaboration\IEEEauthorrefmark{2}}
                            \\
\IEEEauthorblockA{\IEEEauthorrefmark{1}Laboratoire AstroParticle and Cosmology \\ 10, rue A. Domon et L. Duquet,
75205 Paris Cedex 13, France.}
\IEEEauthorblockA{\IEEEauthorrefmark{2}http://antares.in2p3.fr}
}

\shorttitle{Baret \etal ANTARES Charge calibration}
\maketitle

\begin{abstract}
ANTARES is a deep-sea, large volume mediterranean neutrino telescope installed off the Coast of Toulon, France. It is taking data in its complete configuration since May 2008 with nearly 900 photomultipliers installed on 12 lines. It is today the largest high energy neutrino telescope of the northern hemisphere. The charge calibration and threshold tuning of the photomultipliers and their associated front-end electronics is of primary importance. It indeed enables to translate signal amplitudes into number of photo-electrons which is the relevant information for track and energy reconstruction. It has therefore a strong impact on physics analysis. We will present the performances of the front-end chip, so-called ARS, including the waveform mode of aquisition. The in-laboratory as well as regularly performed in situ calibrations will be presented together with related studies like the time evolution of the gain of photomultipliers

  \end{abstract}

 
\section{Introduction}
ANTARES is an underwater neutrino telescope
installed at a depth of 2475 m in the
Mediterranean Sea. The site is at
about 40 km off the coast of Toulon, France. The
control station is installed in Institut Michel Pacha in La Seyne Sur Mer, close to Toulon.
The apparatus consists of an array of 900 photo-multiplier tubes (PMTs) by which the faint light
pulses emitted by relativistic charged particles 
propagating in the water may be detected. Based on such
measurements, ANTARES is capable of
identifying neutrinos of atmospheric as well as of
astrophysical origin. In addition, the detector 
is a monitoring station for geophysics and
sea science investigations.
For an introduction to the scientific aims of the
ANTARES experiment, the reader is referred to
the dedicated presentation at this Conference \cite{ant_sc}.

\section{The ANTARES apparatus}
                                   
The detector consists of an array of 900 large area photomultipliers (PMTs), Hamamatsu
R7081-20, enclosed in pressure-resistant glass
spheres to constitute the optical modules (OMs)\cite{om}, and arranged on 12 detection lines. An
additional line is equipped with
environmental devices.						
Each line is anchored to the sea bed and kept close to 
vertical position by a top buoy. The minimum	
distance between two lines ranges from 60 to 80~m.
Each detection line is composed by 25 storeys,
each equipped with 3 photomultipliers oriented downward at
$45^{\circ}$ with respect to the vertical. The storeys are
spaced by 14.5 m, the lowest one being located	
about 100 m above the seabed. 	
From the functional point of view, each line is	
divided into 5 sectors, each of which consists	
typically of 5 storeys. Each storey is controlled by
a Local Control Module (LCM), and each sector	
is provided with a modified LCM, the Master	
Local Control Module (MLCM), which controls
the data communications between its sector and	
the shore. A String Control Module (SCM),
located at the basis of each line, interfaces the line	
to the rest of the apparatus.
Each of these modules consists of an aluminum
frame, which holds all electronics boards
connected through a backplane and is enclosed in			
a water-tight titanium cylinder.

\section{The front-end electronics}

 \begin{figure*}[!t]
   \centerline{
	\subfloat{
  	\includegraphics[width=5.3in,height=2.8in]{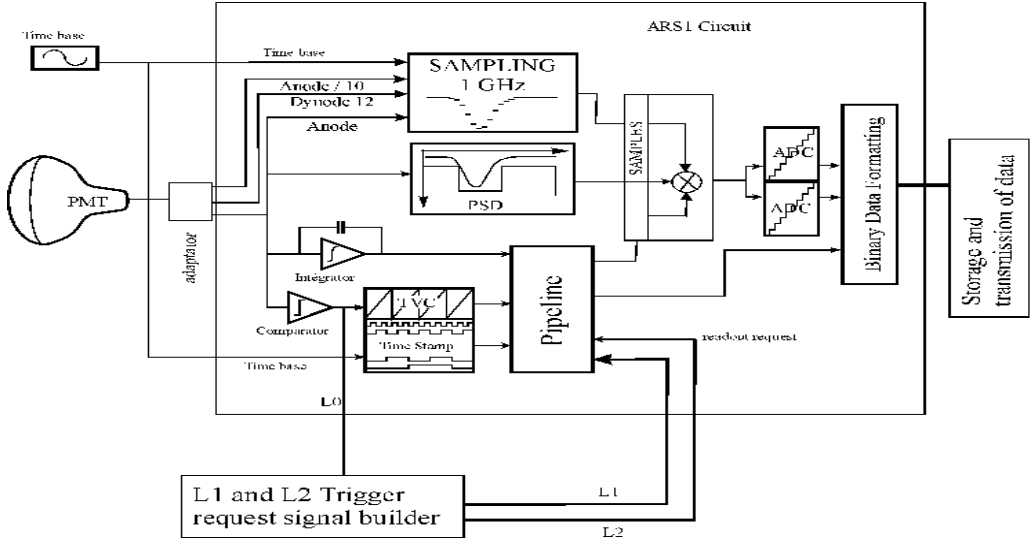}
	}
	}
  \caption{ARS architecture}
  \label{ars_arch}
 \end{figure*}

The full-custom Analogue Ring Sampler (ARS)
has been developed to perform the complex front-end operations \cite{ars}.This chip samples the PMT signal
continuously at a tunable frequency up to 1 GHz
and holds the analogue information on 128
switched capacitors when a threshold level is crossed.
The information is then digitized, in response to a
trigger signal, by means of two integrated dual 8-
bit ADC. Optionally the dynamic range may be
increased by sampling the signal from the last dynode.
A 20 MHz reference clock is used for time
stamping the signals. A Time to Voltage
Converter (TVC) device is used for high-resolution time measurements between clock
pulses. The ARS is also capable of discriminating
between simple pulses due to conversion of single
photoelectrons (SPE) from more complex
waveforms. The criteria used to discriminate
between the two classes are based on the
amplitude of the signal, the time above threshold
and the occurrence of multiple peaks within a
time gate. Only the charge and time information
is recorded for SPE events, while a full waveform
analysis is performed for all other events.
The ARS chips are arranged on a motherboard to
serve the optical modules. Two ARS chips, in a
``token ring'' configuration, perform the charge and
time information of a single PMT. A third chip
provided on each board is used for triggering
purposes.
The settings of each individual chip can be
remotely configured from the shore.

\section{Test bench calibration}
\begin{figure}[!t]
  \centering
\includegraphics[width=2.5in]{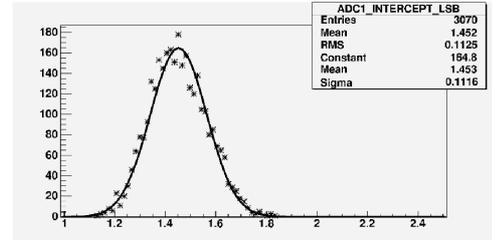} 
  \caption{Distribution of the intercept (in mV) of the ADC transfer function}
\label{avc_intercept}
 \end{figure}

The bare ARSs were calibrated at IRFU-CEA/Saclay. There, the transfer functions of the Amplitude to Voltage Converter (AVC) have been measured. This AVC transfer function is an important parameter for the correction of the walk of the PMT signal and also for measurement of the amplitude of each PMT pulse.
The principal component of this bench is a pulse generator which directly sends signals to a pair of ARSs operating in a flip-flop mode. The generated pulse is a triangle with 4 ns rise time and 14 ns fall, somewhat
similar to the electrical pulse of a PMT with variable amplitude. 
 The tranfer functions of the dynamic range of the ADCs are linear and parametrised by their slope and intercept. The distributions of these two parameters for a large sample of ARS chips are presented on figure \ref{avc_slope} and \ref{avc_intercept} and one can see that they are homogeous which enables to use the same parameters for all ARSs.


 \begin{figure}[!t]
  \centering
  \includegraphics[width=2.5in]{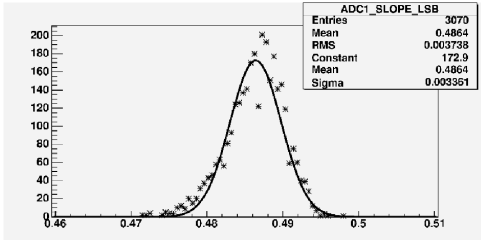} \label{avc_slope}
  \caption{Distribution of the slope (in mV/bit) of the ADC transfer function}
  \label{avc_slope}
 \end{figure}

\section{In Situ calibration}

 \begin{figure}[!t]
  \centering
  \includegraphics[width=2.6in,height=1.9in]{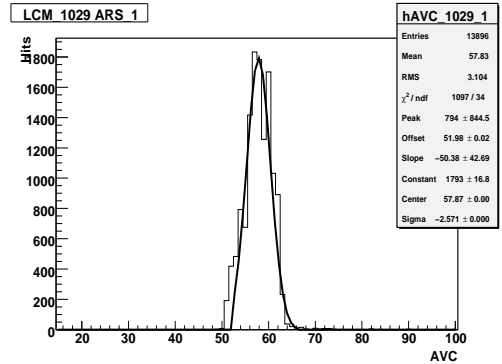}
  \caption{Example of AVC distribution of photoelectron like data with the corresponding fit (see text).}
  \label{1pe}
 \end{figure}

Specials runs reading the PMT current at random times allow to measure the corresponding so-called pedestal value of the AVC channel. Besides, the photoelectron peak can easily be studied with minimum bias events since the optical activity due the ${}^{40}K$ decays and bioluminescent bacteria produces, on average, single photons at the photocathode level. The knowledge of the photoelectron peak and the pedestal is used to estimate the charge over the full dynamical range of the ADC. The integral linearity of the ADC used in the ARS chip has independently been studied using the TVC channel and shows satisfactory results \cite{timecalib}. 
An example of in situ charge distribution for a particular ARS is shown on figure \ref{1pe}.
The values in AVC channel of the pedestal and the photoelectron peak are used to convert individual measurements into photoelectron units. Charge distributions obtained with minimum bias data (based on snapshot of the overall activity of an optical module above a given threshold) can be parameterized using the following simple formula:
\begin{equation}
    Ae^{-\alpha (x-x_{th})}+Be^{\frac{-(x-x_{pe})^2}{2\sigma^2}}
\end{equation}

 \begin{figure}[!t]
  \centering
  \includegraphics[width=2.6in,height=1.9in]{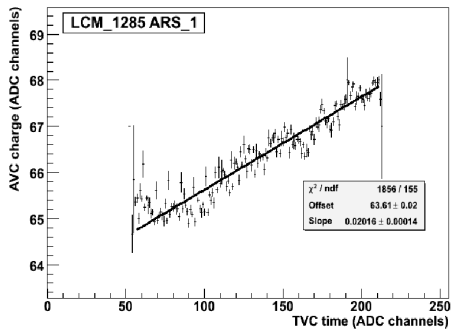}
  \caption{Example of the observed cross talk effect form the time measurement channel affecting the charge measurement channel.}
  \label{Xtalk}
 \end{figure}
\begin{figure}[!t]
  \centering
  \includegraphics[width=2.8in,height=2.in]{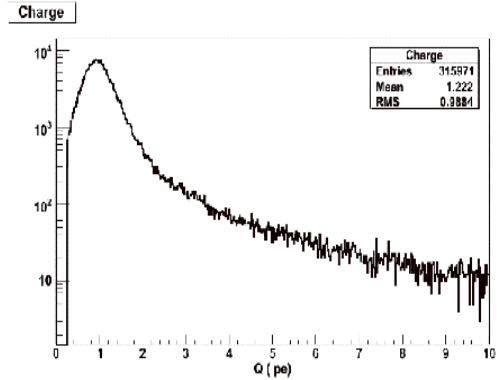}
  \caption{Charge distribution in $pe$ recorded by the detector.}
  \label{min_bias_charge}
 \end{figure}
\begin{figure}[!t]
  \centering
  \includegraphics[width=2.5in,height=2.in]{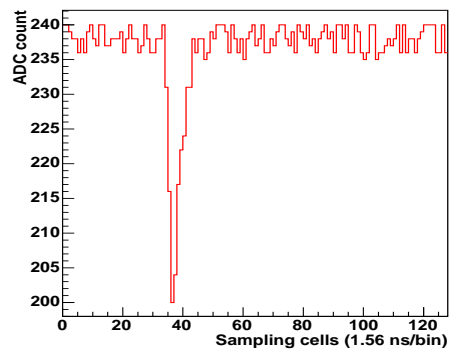}
  \caption{Example of a waveform sampling of an OM signal}
  \label{WF}
 \end{figure}
\begin{figure}[!t]
  \centering
  \includegraphics[width=2.5in,height=2.in]{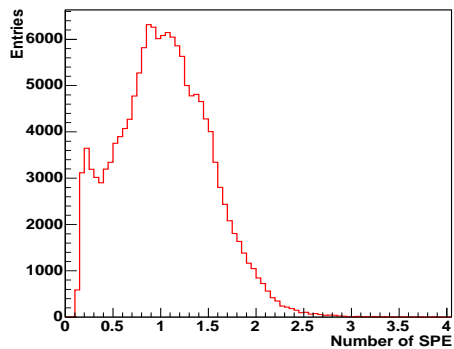}
  \caption{Charge distribution
of the PMT signal obtained by integrating the WF samples }
  \label{QWF}
 \end{figure}

The first term accounts for the dark current of the PMT, the second one describes the photoelectron distribution itself. The parameters $x_{th}$ and $x_{pe}$ are respectively the effective thresholds (``offset'') and photoelectron peak (``center'') in AVC units.\\ 
The charge measurements in the AVC channels appear to be influenced by the time measurements in the TVC channel (the inverse effect does not apply). This effect, referred to as ``cross talk effect'' can be, considering the current settings of the ARS, on an event-by-event basis
 as high as 0.2~pe. It is thought to be due to a cross talk of the capacitors inside the ARS pipeline.  It is a linear effect that does not require correction on high statistics basis (when hits populate the full range of the TVC, the effect washes out). Nevertheless a correction has to be applied to the measured charge of a single event. This correction can be inferred with in situ measurements at the level of the photoelectron by plotting the AVC value against the TVC value as can be seen in figure \ref{Xtalk}. After calibration including cross talk correction, minimum bias events recorded by the detector, coming predominantly from ${}^{40}K$ decay and bioluminescence are dominated by single photo-electron charges as is shown on figure \ref{min_bias_charge}\\
 \\
The ARS has also the capability to perform a
full waveform sampling (WF) of the OM signal in addition to the charge measurement
of the PMT pulse and its arrival time. Although this functionality
is mainly used to record double pulses or large
amplitude signals, it is useful to cross-check the computation of the
SPE charge by the integrator circuit of the ARS. In WF mode, 128
digitisations of the OM anode signal are provided, at a sampling rate
of 640 MHz. In order to obtain a precise time stamping of the WF data,
a synchronous sampling of the 50~MHz internal ARS clock is also
performed and read out in addition to the OM data. An example of a WF
record is shown in figure~\ref{WF}.
Figure~\ref{QWF} 
displays the charge distribution of the OM signals
obtained by integrating the WF samples after baseline subtraction. The
single photo-electron peak is clearly identified well above the
electronics noise.
 
\begin{figure}[!t]
  \centering
  \includegraphics[width=2.5in, height=1.3in]{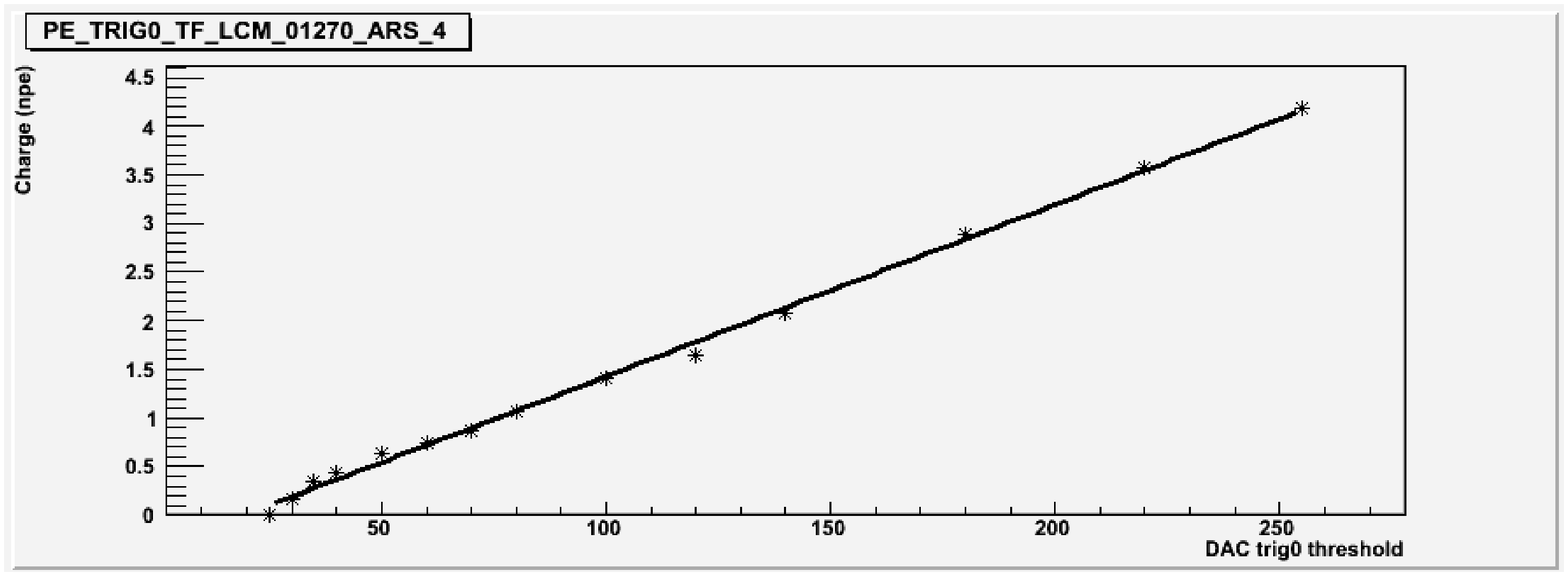}
  \caption{Example of an effective threshold transfer function of an ARS obtained with TS=0 events.}
  \label{thr_insitu}
 \end{figure}

When the time over threshold of a pulse is too short, the ARS chip cannot properly generate the time stamp (TS) of the event, which remains null. This happens when the hit amplitude is just above the amplitude threshold.  The charge (AVC) and fine time of the events (TVC) are recorded correctly. This specific behaviour has negligible influence on efficiency but enables to measure the effective threshold in AVC units for different DAC settings by selecting event with TS=0. For different slow control DAC values the mean AVC values of events at the threshold are recorded during these special calibration runs. The result of the linear fit  of the transfer  function gives the intercept (DAC value for null threshold) and slope. An example is shown on figure \ref{thr_insitu}.
 \\
This method can be applied to every readout ARS of the detector. There are therefore individual in situ calibration and transfer functions for each ARS. These monitored values are stored in a dedicated database and used for further adjustments of the detector setting. In particular, these effective calibrations are used to homogenize the individual thresholds to a value close to 1/3 pe.  

The desintegration of ${}^{40}K$ present in sea water can be used to monitor the evolution with time of the detector response. Indeed, relativistic electrons produced by the $\beta$ desintegration will produce Cerenkov photons which can trigger two adjacent optical modules in the same storey. 
 \begin{figure}[!t]
  \centering
  \includegraphics[width=3.in]{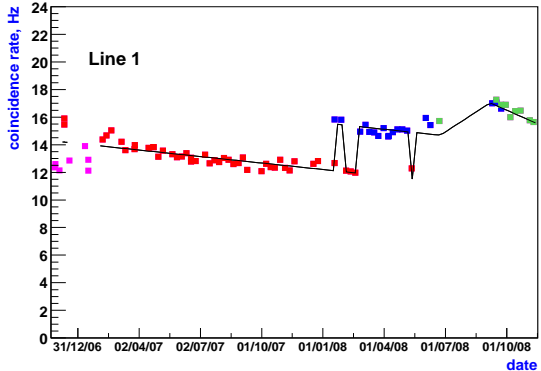}
  \caption{Time evolution of the counting rate due to ${}^{40}K$ desintegration measured with the first line of the detector. Colours account for different threshold setting periods.}
  \label{40K}
 \end{figure}

\begin{figure}[!t]
  \centering
  \includegraphics[width=2.5in]{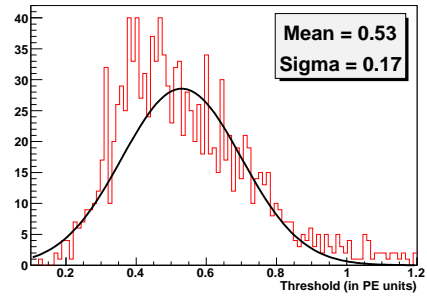}
  \caption{Distributions of the effective thresholds before (12/2007) in situ tuning procedure.}
  \label{th_before}
 \end{figure}

\begin{figure}[!t]
  \centering
  \includegraphics[width=2.5in]{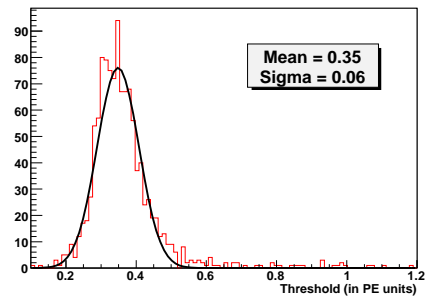}
  \caption{Distributions of the effective thresholds after (03/2008) in situ tuning procedure.}
  \label{th_after}
 \end{figure}

As can be seen on figure \ref{40K}, ${}^{40}K$ counting rate evolution with time have shown a regular decrease. This PMTs gain drop effect is thought to be due to ageing effect of the photocathode. As can be seen after the period between july and september of 2008 when the detector was off for cable repair, gain seems to be partially recovered when PMTs are off for some time.\\
Since all channels are tunned to have an effective threshold of $0.3~\textrm{pe}$, one has to regularly check, and if necessary correct the value of the effective thresholds. This is done thanks to the TS=0 events as explained earlier. The effect of this procedure can be seen on the effective threshold distribution on figure \ref{th_before} and \ref{th_after}.  
 

\section{Conclusions}
All the ARSs of the ANTARES neutrino telescope have been calibrated prior to deployment in order to be able to translate the electrical signal from the chips into number of photo-electrons which is the relevant information for event recontruction and physics analysis. Furthermore, in situ calibration procedures have been developped and are regularly performed in order to monitor and control the detector response, especially to take into account PMTs gain evolution.  

\section{Acknowledgement}
This work was in part supported by the French ANR grant ANR-08-JCJC-0061-01.

\end{document}